\documentclass[twocolumn,superscriptaddress,showpacs,preprintnumbers,amsmath,amssymb,10pt]{revtex4-1}

\usepackage{amsmath}
\usepackage{amssymb}
\usepackage{amsfonts}
\usepackage{graphicx}
\usepackage{morefloats}
\usepackage[usenames,dvipsnames]{xcolor}
\usepackage{blkarray}
\usepackage{verbatim}
\usepackage{hyperref}
\usepackage{subfigure}

\begin{document}
\title{Tackling information asymmetry in networks:\\a new entropy-based ranking index}

\author{Paolo Barucca}
\affiliation{University of Z\"urich, Department of Banking and Finance, Z\"urich, ZH,
Switzerland}
\affiliation{London Institute for Mathematical Sciences, 35a South St, Mayfair, London
W1K 2XF, United Kingdom}
\author{Guido Caldarelli}
\affiliation{IMT School for Advanced Studies, P.zza S.Francesco 19, 55100 Lucca (Italy)}
\affiliation{London Institute for Mathematical Sciences, 35a South St, Mayfair, London
W1K 2XF, United Kingdom}
\author{Tiziano Squartini}
\email{tiziano.squartini@imtlucca.it}
\affiliation{IMT School for Advanced Studies, P.zza S.Francesco 19, 55100 Lucca (Italy)}

\date{\today}

\begin{abstract}
Information is a valuable asset for agents in socio-economic systems, a significant part of the information being entailed into the very network of connections between agents. The different interlinkages patterns that agents establish may, in fact, lead to asymmetries in the knowledge of the network structure; since this entails a different ability of quantifying relevant systemic properties (e.g. the risk of financial contagion in a network of liabilities), agents capable of providing a better estimate of (otherwise) unaccessible network properties, ultimately have a competitive advantage. In this paper, we address for the first time the issue of quantifying the information asymmetry arising from the network topology. To this aim, we define a novel index - InfoRank - intended to measure the quality of the information possessed by each node, computing the Shannon entropy of the ensemble conditioned on the node-specific information. Further, we test the performance of our novel ranking procedure in terms of the reconstruction accuracy of the (unaccessible) network structure and show that it outperforms other popular centrality measures in identifying the ``most informative'' nodes. Finally, we discuss the socio-economic implications of network information asymmetry.
\end{abstract}

\pacs{89.75.-k, 89.75.Fb, 89.70.Cf,64.60.Aq}

\maketitle

\section*{Introduction}

Recognizing the most relevant nodes in a networked system represents a topic of growing interest. This translates into identifying nodes with key features, be they structural or functional. Depending on the system under study, in fact, possessing certain features may translate into accessing a competitive advantage or prominent position in the system. The problem has been tackled by defining a plethora of indices, aiming at quantifying the importance of a node in a given system: the so-called centrality measures \cite{newman1,bloch,borgatti,benzi}. 

The latter are intended to capture the role played by each node within the network by optimizing an opportunely-defined objective function: examples are provided by the degree-centrality (defined by the number of neighbors of each vertex) \cite{bloch}, the closeness-centrality (defined by the average distance of the reachable nodes from any, given, node) \cite{sabidussi}, the PageRank-centrality (defined by the number of ``authoritative'' nodes pointing at the vertex under consideration) \cite{page}, etc.

All these centrality measures look at the topological role that nodes have in a network, disregarding the ability of a node to obtain information about the rest of the system. On the contrary, our methodology focuses on the difference in the information contained in the different pattern of interlinkages of each node. We refer to this difference as to \emph{network information asymmetry} and we will show how it allows nodes to obtain a significantly-better estimation of the (otherwise unaccessible) network properties.

Our novel index measures the reduction of uncertainty over the remaining interlinkages that a node's own ego-network allows: the node whose accessible information provides the largest uncertainty reduction will be identified as being the ``most informative'' one. More quantitatively, for each node the uncertainty reduction is computed by comparing the Shannon entropy benchmark value - measurable by all nodes - with the one obtained by also conditioning on the information ego-network on top of it.

Several attempts to define entropy-based indices have been made \cite{local1,bianconi1,bianconi2,borgatti1}; however, the measures that have been proposed so far are based on specific definitions of Shannon entropy, an evidence that severely affects their applicability. As we will show in what follows, InfoRank can be understood as a generalization of these measures, applicable to any maximum-entropy ensemble and to any subset of nodes.

The paper is organized as follows. In Section 1, we introduce the general methodology for computing InfoRank in any ensemble of networks, given any set of commonly shared information. Then, we apply this general methodology to the case of the configuration model. In Section 2, we measure InfoRank on a number of real-world networks and verify its correlation with the reconstruction accuracy achieved by each node. Finally, in Section 3 we discuss the role of network information asymmetry in social, economic and financial systems.

\section{Methods}

\subsection*{InfoRank: theoretical foundations}

This first subsection is devoted to illustrate the theoretical foundations of our proposed methodology, rooted into information theory. Let us focus on the simplest case of a single node: in order to calculate InfoRank each node (hereafter indexed by $l$) can be imagined to solve two different problems. The first one reads

\begin{eqnarray}\label{entropy1}
S&=&-\sum_\mathbf{G} P(\mathbf{G})\ln P(\mathbf{G})+\nonumber\\
&&-\sum_{i=0}^M\eta_i\left[\sum_\mathbf{G}P(\mathbf{G})C_i(\mathbf{G})-C_i^*\right]
\end{eqnarray}
with $S$ indicating the \emph{constrained} Shannon entropy and $C_0(\mathbf{G})=C_0^*=1$ encoding the normalization condition. By solving the constrained-optimization problem above, node $l$ finds that

\begin{equation}\label{l1}
S=\sum_{i=1}^M\eta_iC_i^*+\ln Z(\vec{\eta}).
\end{equation}
(where $Z(\vec{\eta})=\sum_{\mathbf{G}}e^{-\sum_{i=1}^M\eta_iC_i(\mathbf{G})}$). On the other hand, the second problem node $l$ has to solve reads

\begin{equation}\label{entropy2}
S_{(l)}=S-\sum_{m}\psi_{lm}\left[\sum_\mathbf{G}P(\mathbf{G})a_{lm}(\mathbf{G})-a^*_{lm}\right]
\end{equation}
with $S_{(l)}$ being nothing else than the functional in \ref{entropy1} further constrained by imposing the node $l$ specific pattern of interconnections (i.e. $a^*_{lm}=0,1$ where $m$ runs over all the other nodes of the network). Upon solving the second constrained-optimization problem, the expression

\begin{equation}
S_{(l)}=\sum_{i=1}^M\theta_iC_i^*+\sum_m\psi_{lm}a^*_{lm}+\ln Z'(\vec{\theta},\vec{\psi})
\end{equation}
(where $Z'(\vec{\theta},\vec{\psi})=\sum_{\mathbf{G}}e^{-\sum_{i=1}^M\theta_iC_i(\mathbf{G})-\sum_m\psi_{lm}a_{lm}(\mathbf{G})}$) is found. Both functionals achieve a minimum in their stationary point (consistently, since we are trying to minimize each node - residual - uncertainty). This can be easily proven, upon noticing that the Hessian matrix of both $S$ and $S_{(l)}$ is the covariance matrix of the constraints and, as such, positive-semidefinite.

In order to find the stationary point of $S_{(l)}$, node $l$ must solve the equations $\frac{\delta S_{(l)}}{\delta \psi_{lm}}=0,\:\forall\:m$, i.e.

\begin{equation}\label{maxlik}
\sum_{\mathbf{G}}\left(\frac{e^{-\sum_{i=1}^M\theta_iC_i(\mathbf{G})-\sum_m\psi_{lm}a_{lm}(\mathbf{G})}}{\mathcal{Z}'(\vec{\theta},\vec{\psi})}\right)a_{lm}(\mathbf{G})=a^*_{lm}
\end{equation}
for each of the $m$ values. In order to numerically evaluate the parameters $\psi_{lm},\:\forall\:m$, let us focus on a specific value, e.g. $\psi_{l1}$; we can then divide $Z'(\vec{\theta},\vec{\psi})$ into two subsums: one where $a_{l1}=0$ and one where $a_{l1}=1$. Thus, condition \ref{maxlik} can be rewritten as

\begin{equation}
\sum_{\mathbf{G}_1}\left(\frac{e^{-\sum_{i=1}^M\theta_iC_i(\mathbf{G}_1)-\psi_{l1}-\sum_{m(\neq 1)}\psi_{lm}a_{lm}}}{Z'(\vec{\theta},\vec{\psi})}\right)=a^*_{l1}
\end{equation}
where the sum runs over the network configurations having $a_{l1}=1$ (and briefly indicated with the symbol $\mathbf{G}_1$). Analogously, $Z'(\vec{\theta},\vec{\psi})=Z_0'(\vec{\theta},\vec{\psi})+e^{-\psi_{l1}}Z_1'(\vec{\theta},\vec{\psi})$ where 

\begin{equation}
Z_0'(\vec{\theta},\vec{\psi})=\sum_{\mathbf{G}_0}e^{-\sum_{i=1}^M\theta_iC_i(\mathbf{G}_0)-\sum_{m(\neq 1)}\psi_{lm}a_{lm}}
\end{equation}
and
\begin{equation}
Z_1'(\vec{\theta},\vec{\psi})=\sum_{\mathbf{G}_1}e^{-\sum_{i=1}^M\theta_iC_i(\mathbf{G}_1)-\psi_{l1}-\sum_{m(\neq 1)}\psi_{lm}a_{lm}},
\end{equation}
the first sum runs over the networks having $a_{l1}=0$ and the second sum runs over the networks having $a_{l1}=1$. Solving the likelihood equation in the case $a^*_{l1}=0$ brings to $\psi_{l1}=+\infty$; solving the likelihood equation in the case $a^*_{l1}=1$ brings to $\psi_{l1}=-\infty$. Thus, if $a^*_{l1}=0$ one has $S_{(l)}=\sum_{i=1}^M\theta_iC_i^*+\ln Z_0'(\vec{\theta},\vec{\psi})$ since the term $Z_1'(\vec{\theta},\vec{\psi})$ is suppressed by the coefficient $e^{-\psi_{l1}}$ which converges to zero; otherwise, if $a^*_{l1}=1$ then $\mathcal{S}_{(l)}=\sum_{i=1}^M\theta_iC_i^*+\ln Z_1'(\vec{\theta},\vec{\psi})$ since the term $Z_0'(\vec{\theta},\vec{\psi})$ is suppressed by the coefficient $e^{\psi_{l1}}$ which converges to zero.

The estimation of the other parameters proceeds in an analogous way, by applying the same line of reasoning to the ``surviving'' partition functions. In other words, specifying the single patterns of connections means reducing the number of configurations over which the estimation of the constraints is carried out: thus, $Z'(\vec{\theta})$ runs over a smaller number of configurations than $Z(\vec{\eta})$.

Let us now evaluate the expressions $Z(\vec{\eta})$ and $Z'(\vec{\theta})$ for the \emph{same value} of the parameters (say $\vec{\mu}$): since the number of addenda in $Z(\vec{\mu})$ is larger than the number of addenda in $Z'(\vec{\mu})$, it also holds true that $\ln Z(\vec{\mu})\geq\ln Z'(\vec{\mu})$ and the inequivalence $S(\vec{\mu})\geq S_{(l)}(\vec{\mu})$ is true as well. Let us now choose a particular value of the parameters, i.e. the point of minimum of $S$: $\vec{\mu}=\vec{\eta}^*$. Thus,

\begin{equation}
S(\vec{\eta}^*)\geq S_{(l)}(\vec{\eta}^*)\geq S_{(l)}(\vec{\theta}^*)
\end{equation}
where the second inequality follows from the very definition of minimum. Our ranking procedure builds upon the evidence that, by imposing more information on top of the common one, each node further reduces its uncertainty about the unknown network structure: the one reducing the residual uncertainty to the largest extent is identified as the ``most informative'' one.
The same reasoning and calculations apply for the case of subsets of nodes, only quantifying the InfoRank of all possible subsets of $s$ nodes in a network of size $N$ would require computing ${N}\choose{s}$ Shannon entropies. 

The next subsections will be devoted to illustrate the steps defining our approach, by focusing, for the sake of simplicity, on (the simplest case of) binary, undirected networks.

\subsection*{Quantifying the benchmark information}

Let us suppose the benchmark information, which is accessible to all nodes in the network $\mathbf{A}$, to be represented by the degree sequence. The benchmark model is, thus, represented by the usual Configuration Model (CM) \cite{newman2}, defined by the following system of equations:

\begin{equation}\label{sysint}
k_i(\mathbf{A})=\sum_{j(\neq i)}\frac{x_ix_j}{1+x_ix_j}\equiv\sum_{j(\neq i)}p_{ij},\:\forall\:i.
\end{equation}

The informativeness of the degree sequence in explaining the network structure can be quantified by calculating the value of the Shannon entropy defined by the chosen constraints:

\begin{eqnarray}\label{entya}
S_0&=&\frac{1}{2}\sum_iS_0^{(i)}=\nonumber\\
&=&-\frac{1}{2}\sum_i\sum_{j(\neq i)}[p_{ij}\ln p_{ij}+(1-p_{ij})\ln(1-p_{ij})]\nonumber\\
\end{eqnarray}
with $S_0^{(i)}$ indicating the contribution of node $i$ to the benchmark entropy $S_0$ (the subscript $0$ stresses the benchmark-like value of this functional, encoding a kind of information which is accessible to all nodes). Intuitively, the closer the $S_0$ value to zero, the larger the explanatory power of the degree sequence with respect to the network structure.

\subsection*{Quantifying the node-specific information}

Let us now focus on a specific node, e.g. $i$. Constraining the information encoded into its specific pattern of connections implies letting node $i$ impose

\begin{equation}
p_{ij}=\frac{x_ix_j}{1+x_ix_j}=a_{ij},\:\forall\:j
\end{equation}
i.e. treating as deterministic the links constituting its ego-network. As an example, let us suppose that node $i$ is linked only with nodes 2 and 3 out of the $N$ constituting our ideal network, i.e. that $\frac{x_ix_2}{1+x_ix_2}=\frac{x_ix_3}{1+x_ix_3}=1$ and $\frac{x_ix_1}{1+x_ix_1}=\frac{x_ix_4}{1+x_ix_4}=\dots=\frac{x_ix_N}{1+x_ix_N}=0$. This implies that the system of equations that node $i$ has to solve becomes

\begin{eqnarray}\label{sys}
k_1(\mathbf{A})=\frac{x_1x_2}{1+x_1x_2}+\dots\:&0&\:\dots+\frac{x_1x_N}{1+x_1x_N}\nonumber\\
k_2(\mathbf{A})=\frac{x_2x_1}{1+x_2x_1}+\dots\:&1&\:\dots+\frac{x_2x_N}{1+x_2x_N}\nonumber\\
k_3(\mathbf{A})=\frac{x_3x_1}{1+x_3x_1}+\dots\:&1&\:\dots+\frac{x_3x_N}{1+x_3x_N}\nonumber\\
k_4(\mathbf{A})=\frac{x_4x_1}{1+x_4x_1}+\dots\:&0&\:\dots+\frac{x_4x_N}{1+x_4x_N}\nonumber\\
&\vdots&
\end{eqnarray}
where we have explicitly specified the value of the coefficients quantifying the probability of node $i$ to establish a connection with every other node (notice that we have omitted the equation controlling for the value of the $i$-th degree, since trivially satisfied). The system above can be rearranged by moving at the left hand side the known entries of the adjacency matrix:

\begin{equation}
k_l(\mathbf{A})-a_{il}=\sum_{j(\neq l,i)}\frac{x_lx_j}{1+x_lx_j}\equiv\sum_{j(\neq l,i)}\tilde{p}_{lj}^{(i)},\:\forall\:l(\neq i)
\end{equation}
i.e. the zeros and the ones characterizing the (missing) links with node $i$. More explicitly:

\begin{eqnarray}
k_1(\mathbf{A})&=&\sum_{j(\neq 1,i)}\frac{x_1x_j}{1+x_1x_j}\equiv\sum_{j(\neq 1,i)}\tilde{p}_{1j}^{(i)}\nonumber\\
k_2(\mathbf{A})-1&=&\sum_{j(\neq 2,i)}\frac{x_2x_j}{1+x_2x_j}\equiv\sum_{j(\neq 2,i)}\tilde{p}_{2j}^{(i)}\nonumber\\
k_3(\mathbf{A})-1&=&\sum_{j(\neq 3,i)}\frac{x_3x_j}{1+x_3x_j}\equiv\sum_{j(\neq 3,i)}\tilde{p}_{3j}^{(i)}\nonumber\\
k_4(\mathbf{A})&=&\sum_{j(\neq 4,i)}\frac{x_4x_j}{1+x_4x_j}\equiv\sum_{j(\neq 4,i)}\tilde{p}_{4j}^{(i)}\nonumber\\
&\vdots&
\end{eqnarray}
where the superscript $(i)$ stresses that the numerical value of the probability coefficients $\{\tilde{p}_{jk}^{(i)}\}$ is induced by the specification of node $i$ patterns of connections and, in general, $\tilde{p}_{jk}^{(i)}\neq p_{jk}$. Notice, in fact, that the problem of quantifying the informativeness of the ego-network of each node can be restated by imagining that the node itself is \emph{removed} from the network, in such a way that a reduced adjacency matrix $\mathbf{\tilde{A}}$ remains naturally defined, inducing, in turn, a reduced system of equations:

\begin{equation}\label{sysred}
\tilde{k}_l(\mathbf{\tilde{A}})=\sum_{j(\neq l,i)}\frac{x_lx_j}{1+x_lx_j},\:\forall\:l(\neq i).
\end{equation}

\subsection*{Calculating the node-specific InfoRank}

Once $i$ has been removed from the network, the entropy of the ``surviving'' topological structure can be computed by employing the novel probability coefficients defined by the system of equations in \ref{sysred}, i.e.

\begin{equation}\label{entyb}
S_{(i)}=-\frac{1}{2}\sum_j\sum_{k(\neq j)}[\tilde{p}_{jk}^{(i)}\ln \tilde{p}_{jk}^{(i)}+(1-\tilde{p}_{jk}^{(i)})\ln(1-\tilde{p}_{jk}^{(i)})].
\end{equation}

Since removing different nodes will, in general, impact on the benchmark entropy $S_0$ differently, a ranking is naturally induced by the amount of ``uncertainty reduction'' caused by the removal of each node. Since our aim is identifying the node(s) possessing the largest amount of information, in order to define a novel ranking index, let us divide $S_{(i)}$ by $S_0$ and take the complement to 1:

\begin{equation}
I_i=1-\frac{S_{(i)}}{S_0};
\end{equation}
as apparent from the definition, the larger the entropy reduction, the higher the rank of the node causing it. In what follows, the index $I_i$ will be referred to as to the InfoRank index.

\subsection*{Approximating the node-specific InfoRank}

Although formally similar, the quantities $S_0^{(i)}$ and $S_{(i)}$, respectively defined in eq. \ref{entya} and eq. \ref{entyb}, are conceptually very different and must not be confused. In fact, while $S_0^{(i)}$ just represents the contribution of node $i$ to the benchmark entropy $S_0$, the second index $S_{(i)}$ quantifies the residual uncertainty after the removal of node $i$, accounting, in particular, for the effect that removing such a node has on the remaining vertices. Whenever such an effect can be ignored (i.e. when diminishing the nodes degree by one unit doesn't affect much the magnitude of the surviving probability coefficients), $S_{(i)}$ can be indeed approximated by $S_0-S_0^{(i)}$, further implying that $I_i\simeq S_0^{(i)}/S_0$. 

Remarkably, the term $S_0^{(i)}$ can be simplified as well. For what concerns our analysis, two cases are worth to be mentioned. The first one concerns sparse networks: when assuming the probability coefficients to satisfy the requirement $p_{ij}\ll1,\:\forall\:i\neq j$, then $p_{ij}\simeq x_ix_j,\:\forall\:i\neq j$, implying that $S_0^{(i)}\simeq-\sum_{j(\neq i)}[p_{ij}\ln p_{ij}-p_{ij}]=-k_i\ln\left(\frac{k_i}{\sqrt{2L}}\right)+k_i$. The second approximation is valid whenever the probability coefficients controlling for the connections of node $i$ are well represented by their average value, i.e. $p_{ij}\simeq\frac{\sum_{j(\neq i)}p_{ij}}{N-1}=\frac{k_i}{N-1}\equiv\overline{p}_{ij}$; in this case, one obtains that $S_0^{(i)}\simeq-(N-1)\left[\overline{p}_{ij}\ln\overline{p}_{ij}+\left(1-\overline{p}_{ij}\right)\ln\left(1-\overline{p}_{ij}\right)\right]$.

\begin{figure}[t!]
\begin{center}
\includegraphics[width=0.45\textwidth]{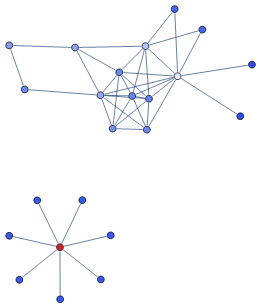}
\caption{Toy network, whose nodes have been ranked according to InfoRank (red nodes are ranked higher than blue nodes). Since the node with the largest score is the one maximally reducing the residual uncertainty of the network, InfoRank is not completely determined by the nodes degrees: the center of the star, in fact, has exactly the same number of neighbors of other nodes (i.e. 7); differently from them, however, its removal would cause and entire portion of the network to become deterministic.}
\label{fig0}
\end{center}
\end{figure}

\section{Results}

\subsection*{Ranking nodes in synthetic networks}

In order to better illustrate the meaning of the InfoRank index, let us consider two extreme cases, i.e. the removal of either an isolated or a fully-connected node. It is intuitive that, in both cases, the knowledge of the connections of the considered nodes adds no information or, equivalently, that removing these nodes doesn't lead to any uncertainty reduction. This is readily seen by comparing the systems \ref{sysint} and \ref{sysred}. In presence of a hub, in fact, the system of equations $k_i(\mathbf{A})=\sum_{j(\neq i)}\frac{x_ix_j}{1+x_ix_j},\:\forall\:i$ can be rewritten as $\tilde{k}_i(\mathbf{A})+1=\sum_{j(\neq i,h)}\frac{x_ix_j}{1+x_ix_j}+1,\:\forall\:i(\neq h)$ (with $h$ denoting the hub). Since solving the latter system with respect to $\{x_i\}_{i\neq h}$ is equivalent to solve the former system, removing a hub doesn't change the information content of the network configuration; analogously, when considering an isolated node. On the other hand, the value $I_i=1$ characterizes a node whose removal induces a configuration which is perfectly deterministic (i.e. composed by isolated nodes, cliques or both). Naturally, in the very special case of a star graph, the central node is both the hub and the vertex with largest InfoRank value.

A relationship between InfoRank and the node degree, nonetheless, exists. In order to understand it, let us start by considering the quantity $S_0^{(i)}=-\sum_{j(\neq i)}[p_{ij}\ln p_{ij}+(1-p_{ij})\ln(1-p_{ij})]$, which corresponds to the Shannon entropy of the $N-1$ possible connections of node $i$: the node bringing the largest contribution to $S_0$, then, is the one maximizing the aforementioned sum, i.e. the one encoding $(N-1)\ln2$ nats into its connections (this unit of measure, also known as ``natural bit'', is a consequence of having chosen the base of the logarithm to be the natural one - in base 2, the overall contribution would have been of $N-1$ bits). Each of the addenda can, thus, be imagined to contribute with an average coefficient $\overline{p}_{ij}=\frac{\sum_{j(\neq i)}p_{ij}}{N-1}=\frac{k_i}{N-1}\simeq\frac{1}{2}$, further implying that a ranking based on the na\"ive contribution of each node to $S_0$ would privilege nodes with $k_i\simeq(N-1)/2$ neighbors.

InfoRank, instead, accounts also for the effect that constraining the pattern of connections of a given node has on the connections of the neighboring ones. Let us consider the synthetic network shown in fig. \ref{fig0}. Upon computing the vector $\{S_0^{(i)}\}$, one finds that the largest contribution to $S_0$ comes from the hub, consistently with the previous discussion (notice, in fact, that $k_h=10\simeq(N-1)/2=11$). Let us know imagine, instead, to remove the center of the star: this would cause, in turn, an entire portion of the network to become deterministically determinable (7 nodes would become isolated, in fact). As a consequence, while the contribution $S_0^{(c)}$ would be enriched by an additional amount of $\simeq-k_c(N-1)\left[p_{leaf}\ln p_{leaf}+\left(1-p_{leaf}\right)\ln\left(1-p_{leaf}\right)\right]$ (with $p_{leaf}=\frac{1}{N-1}$), removing the hub would just disconnect two more nodes (by retaining only the main contribution). InfoRank correctly assigns the highest score to the center of the star, pointing it out as the vertex establishing the most informative set of interconnections. Our index, in other words, encodes higher-order corrections to the na\"ive contribution $S_0^{(i)}$, by including the ``effects'' of the additional constraints on the neighboring vertices.

\begin{figure}[t!]
\begin{center}
\includegraphics[width=0.39\textwidth]{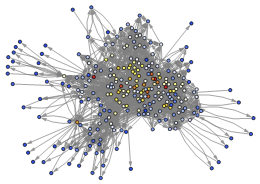}
\includegraphics[width=0.39\textwidth]{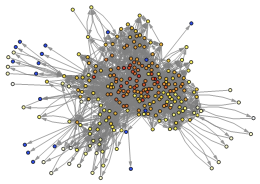}
\includegraphics[width=0.39\textwidth]{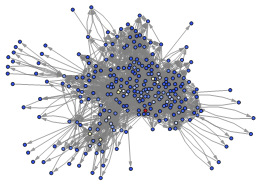}
\includegraphics[width=0.39\textwidth]{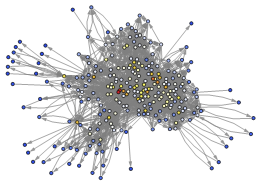}
\caption{\emph{C. Elegans} neural network \cite{tokyo}. From top to bottom, nodes are ranked according to their (out-)degree-centrality, closeness-centrality, PageRank-centrality, InfoRank (red nodes are ranked higher than blue nodes). Notice that, according to PageRank, (only) the node with largest in-degree is ranked first; the same node, however, is characterized by a zero out-degree which, in turn, causes its closeness-induced score to be zero as well.}
\label{fig1}
\end{center}
\end{figure}

\begin{figure*}[t!]
\begin{center}
\includegraphics[width=0.32\textwidth]{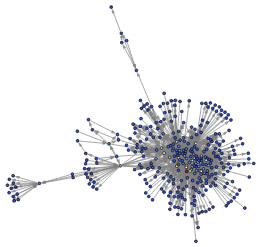}
\includegraphics[width=0.32\textwidth]{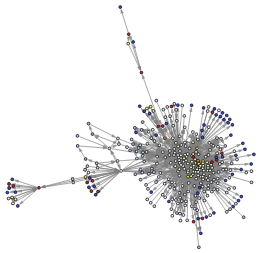}
\includegraphics[width=0.32\textwidth]{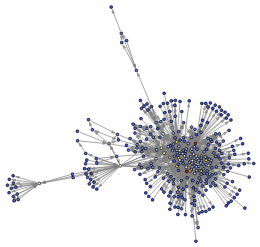}
\caption{US airports network in 1997 \cite{colizza}. Nodes are ranked according to their value of (out-)degree-centrality (left), closeness-centrality (center) and InfoRank (right - red nodes are ranked higher than blue nodes). Nodes with a large (out-)degree-centrality (hubs) do not necessarily coincide with the nodes characterized by a large value of closeness-centrality: in fact, although many nodes can be reached by a walker leaving the hubs, these may lie at a large distance from it.}
\label{fig2}
\end{center}
\end{figure*}

\subsection*{Ranking nodes in real-world networks}

Let us now employ InfoRank to analyse real-world configurations. The core of our analysis will consist in a thorough comparison of a number of alternative ranking indices (in what follows, binary, \emph{directed} networks will be considered, since one of the chosen indicators becomes trivial in the undirected case): in order to consistently compare the ranking scores output by the selected algorithms, the former ones are normalized in order to let them range within the same interval. More specifically, if we let $R_i^{(a)}$ represent the rank of node $i$ according to the chosen algorithm $a$, the applied transformation reads $\overline{R}_i^{(a)}=(R_i^{(a)}-\min\{R_i^{(a)}\})/(\max\{R_i^{(a)}\}-\min\{R_i^{(a)}\})\in[0,1]$ and ensures that nodes with minimum rank are assigned a value $\overline{R}_i^{(a)}=0$ (in blue, according to the color scale adopted throughout the paper); viceversa, nodes with maximum rank are assigned a value $\overline{R}_i^{(a)}=1$ (in red, according to the color scale adopted throughout the paper).

\begin{figure}[t!]
\begin{center}
\includegraphics[width=0.41\textwidth]{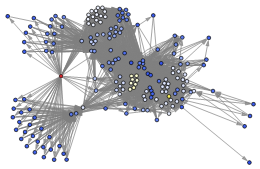}
\includegraphics[width=0.41\textwidth]{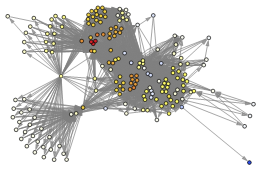}
\includegraphics[width=0.41\textwidth]{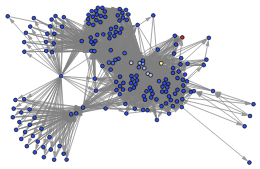}
\includegraphics[width=0.41\textwidth]{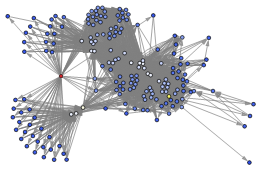}
\caption{Little Rock food web \cite{martinez}. From top to bottom, nodes are ranked according to their (out-)degree-centrality, closeness-centrality, PageRank-centrality, InfoRank (red nodes are ranked higher than blue nodes). Nodes with large in-degree do not necessarily coincide with nodes having a large value of PageRank: this is evident in the case of food-webs, where species exist that are predated by a limited number of predators of which constitute the only preys. In other networks however, the correlation between the PageRank value and the in-degree is quite large.}
\label{fig3}
\end{center}
\end{figure}

The first alternative index is represented by the \emph{degree-centrality}, identifying the nodes characterized by the largest degree as the most important (i.e. central) ones. A first limitation of such an index lies in the nature of the connectivity concept, which lacks an obvious generalization to, e.g. the directed case we are considering in the present paper. In what follows we will adopt the following definition 

\begin{equation}
D_i=k_i^{out}
\end{equation}
which ranks the nodes according to the number of their out-neighbors. As evident from the first panel of fig. \ref{fig1}, fig. \ref{fig2} and fig. \ref{fig3}, the (out-)degree-centrality trivially identifies the hubs as the most central nodes. 

The second indicator we have considered is the so-called \emph{closeness-centrality} \cite{sabidussi}, defined as 

\begin{equation}\label{close}
C_i=\frac{1}{\overline{d}_i}=\frac{\kappa_i}{\sum_jd_{ij}}
\end{equation}
i.e. as the reciprocal of the average topological distance of a vertex from the other, connected ones ($\kappa_i$ is the number of nodes that can be reached from $i$ - following the links direction - and $d_{ij}$ is the topological distance separating $i$ from any reachable node $j$). Intuitively, any two nodes are said to be ``close'' when their distance is ``small'', i.e. few links must be walked to reach one from the other. Naturally, the nodes with $C_i=0$ are the ones with zero out-degree, while a node with exactly $N-1$ connections will be also the most central one. Generally speaking, however, nodes with small degree do not necessarily have a small closeness-centrality value: an example is provided by the second panel of fig. \ref{fig2}, where nodes behaving like ``local hubs'' (e.g. at the center of star-like subgraphs) are, in fact, characterized by a large $C_i$ independently from their degree. On the other hand, nodes with a large degree do not necessarily have a large closeness-centrality value: in fact, the first panel of fig. \ref{fig3} shows that although a large number of nodes can be reached from the hub, many lie at a large distance from it. 

Interestingly, as the second panels of fig. \ref{fig1} and fig. \ref{fig3} show, the nodes minimizing the (average) topological distance from them are the ones connected to a strongly connected component (SCC - either belonging to it or not). In the case of \emph{C. Elegans} neural network, its large reciprocity ($\simeq 0.43$) further levels out the differences between the $C_i$ values of such vertices.

The third indicator considered in the present analysis is the \emph{PageRank-centrality} \cite{page}. It is computed by solving to iterative equation

\begin{equation}
P_i=\frac{1-\alpha}{N}+\alpha\sum_j\left(\frac{a_{ji}}{k_j^{out}}\right)P_j
\end{equation}
which can be imagined to describe a Markov chain: if $a_{ji}=1$ a walker moves from $j$ to $i$ with probability $\frac{1-\alpha}{N}+\frac{\alpha}{k_j^{out}}$; if $a_{ji}=0$ such a probability becomes $\frac{1-\alpha}{N}$ (in a sense, the walker ``jumps'' from $j$ to $i$). The introduction of the addendum accounting for jumps guarantees the convergence of the formula above to the stationary distribution of this dynamical process (which exists and is unique - its Markov chain, in fact, becomes strongly connected and aperiodic by construction) which also provides the searched ranking scores. In what follows we have set $\alpha=0.85$.

By oversimplifying a bit, PageRank scores higher nodes that are pointed either 1) by a large number of nodes which, in turn, have low out-degree (thus becoming ``authoritative'' nodes) or 2) by authoritative nodes themselves \cite{page}. The evidence that nodes with a large PageRank value do not necessarily coincide with the nodes having a large in-degree is provided by the Little Rock food web: in this particular case, a couple of species predated by a limited number of predators can be, indeed, observed; the former, however, constitute the only preys of the latter. In all the other cases the correlation coefficient between the vectors $\{P_i\}$ and $\{k_i^{in}\}$ is quite large: 0.70 for the US airports network (in 1997), 0.82 for the \emph{C. Elegans} neural network, $0.99$ for both the World Trade Web and the e-MID interbank network (notice that upon lowering $\alpha$ the two vectors become less correlated, since the random contribution to the dynamics becomes the prevalent one). Such a correlation has been also noticed elsewhere \cite{santo}.

Let us now consider our novel InfoRank index. As a first observation, the ranking induced by it shows a little overlap with the one provided by the other indices, thus confirming its degree of novelty. The intuitive idea according to which the nodes with largest InfoRank are the ones disconnecting the largest number of subgraphs is confirmed upon looking at the fourth panel of fig. \ref{fig1} and the third panel of fig. \ref{fig2}: when considering either the \emph{C. Elegans} neural network or the US airport networks, in fact, vertices acting as ``junctions'' between a group of leaves and the remaining part of the network are often assigned an InfoRank value that is larger than the one assigned to the ``most internal'' nodes. Naturally, when directed networks are considered, reducing uncertainty does not necessarily imply isolating nodes: it is often enough to exactly determine either their out- or in-degree to gain a notable amount of information.

\begin{figure*}[t!]
\begin{center}
\includegraphics[width=0.45\textwidth]{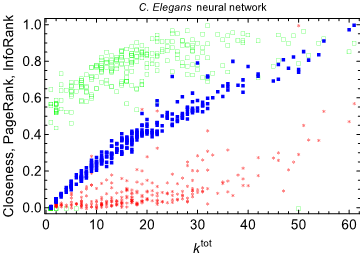}\hspace{1mm}
\includegraphics[width=0.46\textwidth]{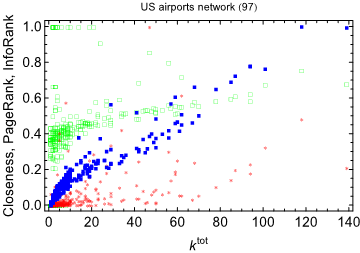}\\
\vspace{2mm}
\includegraphics[width=0.45\textwidth]{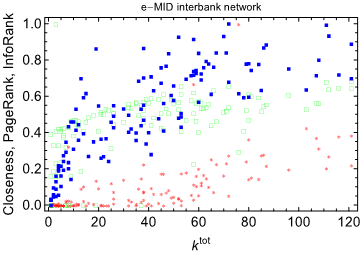}\hspace{1mm}
\includegraphics[width=0.45\textwidth]{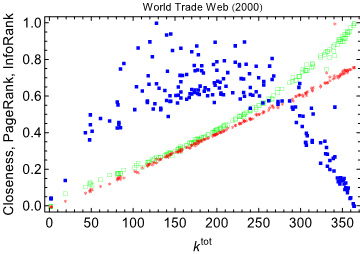}
\caption{Dependence of the (rescaled) ranking indices considered for the present analysis (closeness-centrality $\textcolor{green}{\square}$, PageRank centrality $\textcolor{red}{*}$, InfoRank $\textcolor{blue}{\blacksquare}$) on the nodes total degree. Notice how both e-MID and the World Trade Web are characterized by a strongly positive correlation between the closeness-centrality and the total degree and the PageRank centrality and the total degree; InfoRank, instead, is characterized by a bell-shaped trend for the same systems, whose point-of-maximum lies close to values $k_i^{tot}\simeq(N-1)/2+(N-1)/2=N-1$. Although the nodes providing the largest contribution to the entropy reduction overlap with the ones maximizing $S_0^{(i)}$, this doesn't imply that the removal of a given node has a small impact on the other vertices (see also fig. \ref{figA}). For what concerns sparser systems, instead, InfoRank shows an overall increasing trend while a clear functional dependence between closeness-centrality and total degree and PageRank centrality and total degree is not visible (a weakly positive correlation between closeness-centrality and total degree is, however, present in the \emph{C. Elegans} neural network).}
\label{fig4}
\end{center}
\end{figure*}

\subsection*{Exploring the InfoRank degree-dependence}

Let us now consider the World Trade Web (WTW) \cite{gledi}. The main reason we include it in our analysis is its link density: being much denser than the other networks considered so far, it also allows us to better understand the relationship between InfoRank and the degree sequence(s). 

Since the WTW topological structure can be deduced with great accuracy from the knowledge of its degree-sequence(s) \cite{squartini2}, we may also expect the latter to be correlated with the ranking indices considered for the present analysis. This is indeed the case. As the fourth panel of fig. \ref{fig4} shows, both the closeness-centrality and the PageRank are highly correlated with the total degree (i.e. $k_i^{tot}=k_i^{out}+k_i^{in}$). The monotonic, increasing, relatioship between total degree and closeness-centrality can be straightforwardly explained by noticing that all countries have established a direct connection with the nodes that can be reached by them via some other (indirect) path. This is not true, for example, for the Little Rock food web shown in fig. \ref{fig3}: in that case, the node with largest out-degree is connected to only some of the nodes constituting a strongly connected sub-component; as a consequence, the overall distance from the set of reachable nodes increases. 

\begin{figure*}[t!]
\begin{center}
\includegraphics[width=0.45\textwidth]{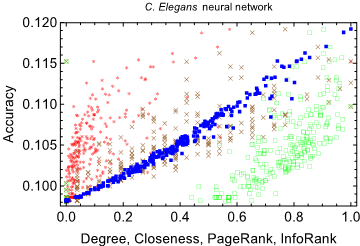}\hspace{1mm}
\includegraphics[width=0.45\textwidth]{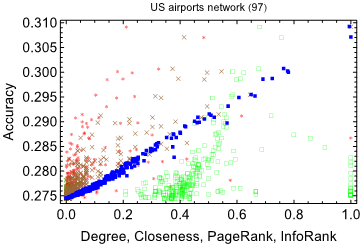}\\
\vspace{2mm}
\includegraphics[width=0.45\textwidth]{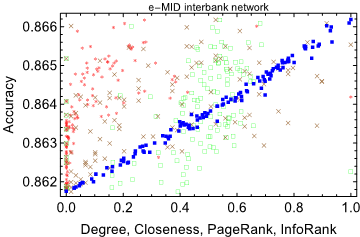}\hspace{1mm}
\includegraphics[width=0.45\textwidth]{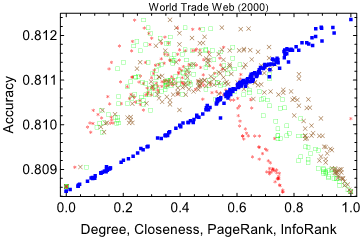}
\caption{Dependence of the accuracy value on the (rescaled) ranking indices considered for the present analysis ((out-)degree-centrality $\textcolor{brown}{\text{x}}$, closeness-centrality $\textcolor{green}{\square}$, PageRank centrality $\textcolor{red}{*}$, InfoRank $\textcolor{blue}{\blacksquare}$). Notice the clear, increasing, trend describing the functional dependence of the accuracy value on the InfoRank value, further confirming that the node(s) establishing the most informative sets of interconnections are the ones characterized by the largest InfoRank value(s).}
\label{fig5}
\end{center}
\end{figure*}

The monotonic, increasing, relationship between total degree and PageRank, instead, rests upon a double (empirical) evidence: countries with a large out-degree are 1) also characterized by a large in-degree and are usually 2) ``pointed'' by countries with a small out-degree.

InfoRank, on the other hand, shows an overall bell-shaped trend with a maximum in correspondence of the values $k_i^{tot}\simeq(N-1)/2+(N-1)/2=N-1$. This means that the nodes providing the largest contribution to the entropy reduction overlap with the ones maximizing $S_0^{(i)}$. However, as evident upon inspecting fig. \ref{figA}, this doesn't mean that the removal of a given node has a small impact on the other vertices; evident deviations from the $S_0^{(i)}$ trend are, in fact, clearly visible: InfoRank adjusts the estimation provided by exclusively accounting for the nodes degrees, although its \emph{functional dependence} on them is, overall, similar to the one characterizing $S_0^{(i)}$.

\subsection*{Exploring the relationship between InfoRank and the reconstruction accuracy}

As we have seen, InfoRank individuates the node(s) reducing the network residual uncertainty to the largest extent. We may, thus, suspect InfoRank to also ``select'' the nodes able to provide the best reconstruction of the network itself. In order to verify our conjecture, we have explicitly tested the agreement between the reconstruction achieved by each node and the observed network structure. In order to do so, we have computed an index often employed to test the (global) goodness of a reconstruction algorithm: the \emph{accuracy}, defined as $\langle A\rangle=\frac{\langle TP\rangle+\langle TN\rangle}{N(N-1)}$ where $\langle TP\rangle$ is the expected number of true positives, i.e. $\langle TP\rangle=\sum_{i}\sum_{j(\neq i)}a_{ij}p_{ij}$, $\langle TN\rangle$ is the expected number of true negatives, i.e. $\langle TN\rangle=\sum_{i}\sum_{j(\neq i)}(1-a_{ij})(1-p_{ij})$ and $N$ is the total number of vertices \cite{squartini1}. We have then summarized our findings by calculating the correlation between the vector $\{A_i\}$ and the vector $\{I_i\}$.

The results are reported in table \ref{tab1}: while the correlation between InfoRank and accuracy is almost 1, when comparing the goodness of the reconstruction achieved by nodes ranked via alternative indices a worse agreement is found. In particular, e-MID and the WTW show a negative correlation value: this is due to the bell-shaped trend recovered, e.g. when scattering the accuracy value versus any of the chosen ranking indicators, consistently with the results illustrated in fig. \ref{fig5}.

\subsection*{Exploring the relationship between InfoRank and systemic risk estimation}

\begin{figure}[t!]
\begin{center}
\includegraphics[width=0.5\textwidth]{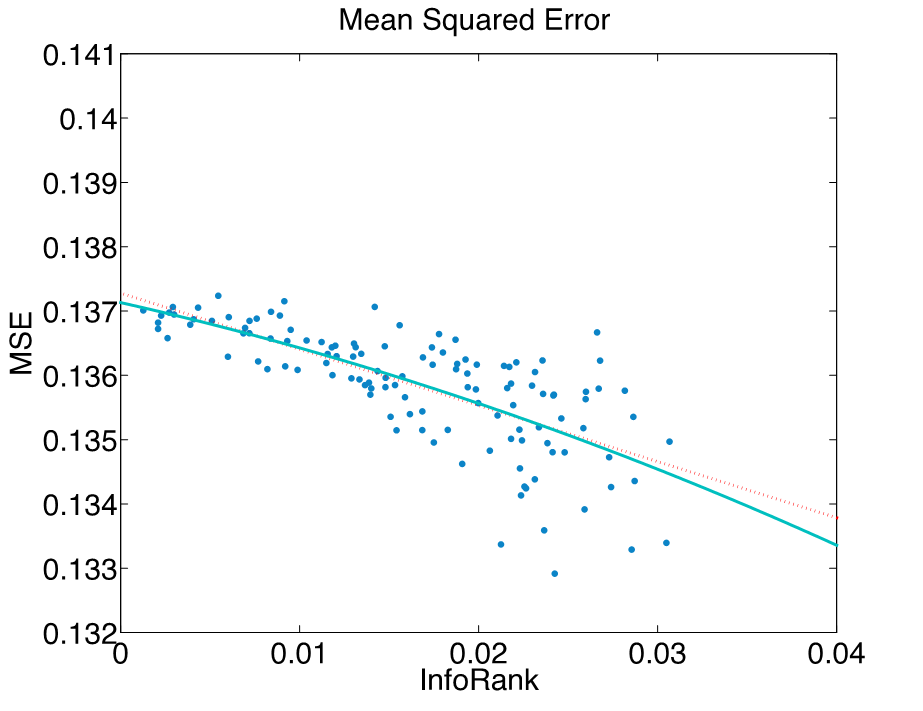}
\caption{Mean squared error over the payment vector of the financial clearing process on the e-MID network. The parameters that account for fire sales effect and insolvency  costs in the Rogers-Veraart \cite{rogers2013failure} clearing mechanism are $\alpha=\beta=.9$. The dotted line is a linear fit of the data $y = - 0.087*x + 0.14$ ($\text{RSS}=0.0070645$), while the solid line is a quadratic fit
$y = - 0.79*x^{2} - 0.063*x + 0.14$ ($\text{RSS}=0.007046$).}
\label{fig6}
\end{center}
\end{figure}

\begin{figure*}[t!]
\begin{center}
\includegraphics[width=0.45\textwidth]{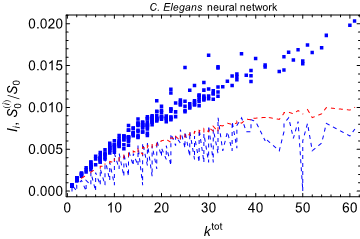}\hspace{1mm}
\includegraphics[width=0.46\textwidth]{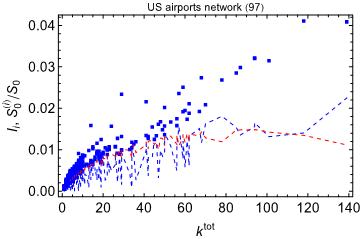}\\
\vspace{2mm}
\includegraphics[width=0.45\textwidth]{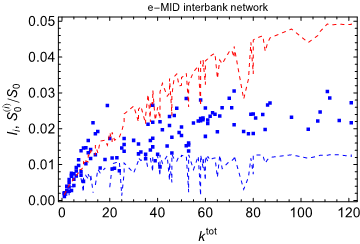}\hspace{1mm}
\includegraphics[width=0.45\textwidth]{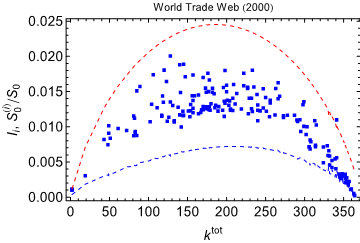}
\caption{Comparison between InfoRank $I_i$ ($\textcolor{blue}{\blacksquare}$) and $S_0^{(i)}$ (blue, dashed line). We have also tested the agreement between the latter and the two approximation derived in the Methods section (red, dashed line). While, for sparse networks, $S_0^{(i)}\simeq-k_i^{out}\ln\left(\frac{k_i^{out}}{\sqrt{L}}\right)+k_i^{out}-k_i^{in}\ln\left(\frac{k_i^{in}}{\sqrt{L}}\right)+k_i^{in}$, for dense networks a non-trascurable difference exists between $S_0^{(i)}$ and $-(N-1)\left[\overline{p}^{out}_{ij}\ln\overline{p}^{out}_{ij}+\left(1-\overline{p}^{out}_{ij}\right)\ln\left(1-\overline{p}^{out}_{ij}\right)\right]-(N-1)\left[\overline{p}^{in}_{ij}\ln\overline{p}^{in}_{ij}+\left(1-\overline{p}^{in}_{ij}\right)\ln\left(1-\overline{p}^{in}_{ij}\right)\right]$ with $\overline{p}^{out}_{ij}=\frac{k_i^{out}}{N-1}$ and $\overline{p}^{in}_{ij}=\frac{k_i^{in}}{N-1}$).}
\label{figA}
\end{center}
\end{figure*}

In this subsection, we try to give a quantitative illustration of how reconstruction accuracy can lead to better estimation of relevant properties of a financial system, focusing on a real network of transaction in an interbank money market. Interbank money markets are essential for financial institutions for the provision of liquidity. In such markets, information asymmetry \cite{wittenberg2008role} on the network of connections can translate in a better estimation of the expected payments, widely recognized as a measure of systemic risk in networks of interbank liabilities \cite{eisenberg}. Here, we focus on data from the e-MID (electronic market of interbank deposit) platform, that served a significant percentage ($\sim 17\%$) of the unsecured money market in the Euro Area before the 2007-2008 crisis \cite{barucca1}.
We apply the clearing mechanism originally proposed in \cite{eisenberg}, in the generalization discussed in \cite{rogers2013failure}, and compute the payment vector, whose components represent the amount a financial institution is able to repay to its creditors. When the payment of a bank is less then its corresponding obligation, that bank is considered insolvent. Hence, computing the payment vector corresponds to identify insolvencies and estimate systemic risk in a financial network. A detailed discussion on such measures of systemic risk is found in \cite{glasserman,barucca2}. Insolvency of bank occurs when its equity, the difference between assets and liabilities, becomes negative. The external cash flow is given by the external assets $A_e$, affected by fire sales in case of insolvency, and external liabilities $L_e$. Both are sampled from a Gaussian distribution, with parameters $\mu_a=10,\, \sigma_a=0.1$ and $\mu_l=1,\, \sigma_l=0.1$, respectively. 
For our analysis on the role of InfoRank, first, we compute the payment vector $\{p_i^{(r)}\}$ - that entails the information on systemic risk losses - starting from the real e-MID network. Secondly, for each node, we compute the payment vector using a sample of networks from its specific ensemble, then evaluate the normalized squared error of each these sampled payment vectors $\{p_i^{(s)}\}$ with respect to the real payment vector. Finally, we calculate the mean over the set of sampled payment vectors and obtain the mean squared error that each node is subjected to in computing this systemic risk measure. 
In Figure \ref{fig6}, we recognize that a larger InfoRank yields a smaller error in the estimation of systemic risk.

\section{Discussion}

We introduced a new index of node relevance in networks based on an information-theoretic approach. 
Differently from other indices, InfoRank can be generalized in several, highly non-trivial, ways. First, its perfectly general derivation allows it to be employed to analyze directed, as well as weighted, networks with any set of commonly shared information. Secondly, it can be extended to quantify the informativeness of whole subsets of nodes: this is usually a major limitation for the other centrality indicators, tailored to provide single nodes estimates. Thirdly, it does not depend on any arbitrary parameter (as PageRank, for example), but only on a maximum entropy principle. 
We verified the financial consequences of the information-theoretic approach we followed in this paper, by applying our algorithm to the estimation of contagion processes in the e-MID network of interbank loans.
To the best of our knowledge, this work is the first to provide a quantitative measure of network information asymmetry leading to competitive advantage for some agents upon others in socio-economic and financial systems.

Finally, the ability of identifying highly informed nodes - characterized by high InfoRank values - may also provide strategies to optimally sample networks, when gathering information on individual nodes is costly, e.g. when surveying a financial system for regulatory purposes.

\begin{table}[t!]
\begin{tabular}{c|c|c|c|c}
\hline
\hline
$r_{A_i,\overline{R}_i}$ & $D_i$ & $C_i$ & $P_i$ & $I_i$ \\
\hline
\hline
Little Rock food web & 0.44 & 0.34 & 0.27 & $\mathbf{0.97}$ \\
\hline
{\it C. Elegans} neural network & 0.82 & 0.60 & 0.65 & $\mathbf{0.98}$ \\
\hline
US airports network (1997) & 0.89 & 0.39 & 0.89 & $\mathbf{0.99}$ \\
\hline
e-MID interbank network & 0.52 & 0.44 & 0.57 & $\mathbf{0.99}$ \\
\hline
World Trade Web (1950) & 0.097 & 0.008 & 0.098 & $\mathbf{0.99}$ \\
\hline
World Trade Web (1970) & -0.1 & -0.23 & -0.15 & $\mathbf{0.99}$ \\
\hline
World Trade Web (2000) & -0.39 & -0.52 & -0.42 & $\mathbf{0.99}$ \\
\hline
\hline
\end{tabular}
\caption{Table showing the Pearson correlation coefficient between the vector of accuracy values $A_i$ and the vector of (rescaled) ranking scores $\overline{R}_i$. The transformation (which doesn't affect the correlation value) reads $\overline{R}_i^{(a)}=(R_i^{(a)}-\min\{R_i^{(a)}\})/(\max\{R_i^{(a)}\}-\min\{R_i^{(a)}\})\in[0,1]$ with $R_i^{(a)}$ representing the rank of node $i$ according to the chosen algorithm $a$.}
\label{tab1}
\end{table}

\section*{Acknowledgements}
PB and TS acknowledge support from: FET Project DOLFINS nr. 640772 and FET IP Project MULTIPLEX nr. 317532.

\end{document}